%% file: main.tex
\documentclass[letterpaper, 10 pt, conference]{ieeeconf}  

\IEEEoverridecommandlockouts                              

\input{additional_packages.tex}

\title{\LARGE \bf Towards a Completeness Argumentation for Scenario Concepts
}

\author{Christoph Glasmacher$^{1}$\orcidlink{0000-0003-4826-9706}, Hendrik Weber$^{2}$\orcidlink{0000-0003-3897-791X}, Lutz Eckstein$^{1}$
\thanks{The work of this paper has been done in the context of the SUNRISE project which is co-funded by the European Commission’s Horizon Europe Research and Innovation Programme under grant agreement number 101069573. The authors want to thank Tino Brade for the fruitful discussions. 
Views and opinions expressed, are those of the author(s) only and do not necessarily reflect those of the European Union or the European Climate, Infrastructure and Environment Executive Agency (CINEA). Neither the European Union nor the granting authority can be held responsible for them.}%
\thanks{Additionally, this research is funded by the VVM project research initiative, promoted by the German Federal Ministry for Economic Affairs and Climate Action (BMWK)}
\thanks{$^{1}$The authors are with Institute for Automotive Engineering RWTH Aachen University, Aachen, Germany {\tt\small name.surname@ika.rwth-aachen.de}}%
\thanks{$^{2}$The author is with RWTH Aachen University, Aachen, Germany }%
}

\begin{document}

\maketitle
\thispagestyle{empty}
\pagestyle{empty}
\copyrightnotice
\vspace{-0.1in}

\begin{abstract}

\input{abstract}

\end{abstract}

\input{body}

\addtolength{\textheight}{-12cm}   

\bibliographystyle{IEEEtran}
\bibliography{bibtex/bib/IEEEabrv, bibtex/bib/paper}

\end{document}

%% file: additional_packages.tex
\usepackage{tabularx}
\usepackage{subfig}
\usepackage{orcidlink}
\usepackage{hyperref}
\usepackage{utfsym}
\usepackage{multirow}
\usepackage{makecell}
\usepackage{booktabs}

\usepackage{layouts}

\usepackage{array}
\newcolumntype{Z}{>{\centering\let\newline\\\arraybackslash\hspace{0pt}}X}

\let\labelindent\relax
\usepackage{enumitem} 
\setlist[itemize]{topsep=3pt, parsep=0pt, leftmargin=1em}

\usepackage{url}

\usepackage{etaremune}

\newcommand\copyrighttext{%
    \footnotesize \copyright{ }2024 IEEE. Personal use of this material is permitted. Permission from IEEE must be obtained for all other uses, in any current or future media, including reprinting/republishing this material for advertising or promotional purposes, creating new collective works, for resale or redistribution to servers or lists, or reuse of any copyrighted component of this work in other works.}
\newcommand\copyrightnotice{%
    \begin{tikzpicture}[remember picture,overlay]
    \node[anchor=south,yshift=15pt,xshift=0pt] at (current page.south) {\parbox{\dimexpr\textwidth-\fboxsep-\fboxrule\relax}{\copyrighttext}};
    \end{tikzpicture}%
}

%% file: abstract.tex
Scenario-based testing has become a promising approach to overcome the complexity of real-world traffic for safety assurance of automated vehicles.
Within scenario-based testing, a system under test is confronted with a set of predefined scenarios. 
This set shall ensure more efficient testing of an automated vehicle operating in an open context compared to real-world testing.
However, the question arises if a scenario catalog can cover the open context sufficiently to allow an argumentation for sufficiently safe driving functions and how this can be proven.
Within this paper, a methodology is proposed to argue a sufficient completeness of a scenario concept using a goal structured notation.
Thereby, the distinction between completeness and coverage is discussed. For both, methods are proposed for a streamlined argumentation and regarding evidence.
These methods are applied to a scenario concept and the inD dataset to prove the usability.

%% file: body.tex
\bstctlcite{MyStyle}

\section{Introduction}
\label{sec:introduction}

Automated driving has become increasingly important in research and the automotive industry. It promises improvements in efficiency and safety of automotive transportation. 
Yet, even though safety benefits are promising, the overall safety of the system needs to be proven by safety assurance processes.
So, safety assurance for automated driving functions (ADF) is a main focus in automotive research. 
Since ADF would have to be tested billions of kilometers using traditional validation methods \cite{Winner2018a}, scenario-based testing has become a promising alternative. 
Thereby, potential traffic situations are split up into dedicated scenarios. 
So, the ADF can be confronted more specifically with pre-defined scenarios instead of facing random situations in the real world.

For this kind of testing, traffic must be subdivided into a countable number of scenarios. 
Because of the complexity of traffic, this is not trivial. 
For this, multiple scenario concepts are developed to describe traffic systematically in a list of distinct scenarios. 
Although multiple scenario concepts are developed, still, the question arises \textit{Is the scenario concept complete? - Or is it at least sufficiently complete?}
Those questions are essential since it has to be ensured that ADS not only work in given scenarios, but they have to work safely in real-world. So, for testing the real-world has to be sufficiently represented by the given set of scenarios.

Within this paper, an argumentative approach is used to propose a methodology to answer those questions. 
Therefore, based on the state of the art, relevant definitions are proposed and the difference between completeness and coverage is elaborated. Afterward, an argumentation framework is set up to tackle the question of completeness of a scenario concept using a goal structured notation (GSN).
This argumentation is exemplarily applied to a given scenario concept \cite{WEB23} and checked with available data to show strategies for both, completeness argumentation and coverage approximation.

\section{Related Work}
\label{sec:related_work}

Although there is no completeness argumentation for scenario concepts to the authors' knowledge, relevant work has been conducted for scenario concepts, completeness approaches, and argumentation techniques.

\subsection{Scenario concepts}
\label{sec:scenario_concepts}

Although there is a common understanding of what a scenario is, there is no universally applicable definition of the term scenario.
\cite{ISO34501} defines a scenario as a sequence of scenes based on \cite{Ulbrich2015}. 
While \cite{Geyer_2014} states that a scene can also cover a short timespan, \cite{Ulbrich2015} defines a scene to be a snapshot of a constellation of traffic participants and their surroundings from the point of view of an omniscient observer.
\cite{Gelder2020a} defines a scenario as a sequence defined by actions and triggers.
According to \cite{ISO34501} a distinction can be made between scenarios, which refer to a single realization of a scenario, and a category that encompasses multiple realizations of scenarios that share a common property.
Furthermore, \cite{ISO34501} defines a distinction between functional, abstract, logical, and concrete scenarios.
\cite{VVM_D13}~refines logical scenarios, differentiating between logical scenario classes, and logical scenario instances.
Where abstract scenarios describe a scenario verbally, logical scenario classes declare parameters for a scenario, and logical scenario instances assign values and distributions.

An important foundation for the generation of scenarios is the 6 layer model, which in its most recent form has been specified by \cite{Scholtes2021} and can also be applied to urban scenarios.
It divides elements within a driving scenario into six layers:
\begin{etaremune}
    \item Digital information
    \item Environmental conditions
    \item Dynamic objects
    \item Temporal modifications of layers 1 and 2
    \item Roadside structures
    \item Road network and traffic guidance objects
\end{etaremune}

Different approaches on how to derive a scenario catalog for verification and validation of automated driving can be identified.
Experts may set up scenario catalogs, based on their experience in system design or testing the system in traffic.
\cite{Guyonvarch2020} is an example of a collection that has been specified for automated vehicles and references an extensive data analysis.
Further relevant collections of scenarios for evaluation of an AV's safety are crash-type catalogs such as the FARS crash type~\cite{NHTSA2022}, or the German three-digit accident type~\cite{GDV2003}, or pre-crash scenarios such as defined by~\cite{Feifel2018}.
A challenge that comes with such expert-based collections is that they are difficult to trace for completeness, as formal information on assumptions and lines of reasoning is not given.

Model-based scenario specification allows the generation of machine-readable scenarios, which can be used for requirements engineering or simulation.
These approaches reference a model, for instance, which maneuvers a traffic participant can execute, to specify the scenarios, e.g. as in~\cite{Bach2016}.
Scenario tagging can be used to define scenarios by means of tags as described by \cite{Gelder2020a}.
These tags have been defined by expert discussion, to cover all relevant aspects for scenario definition.
Tags can be used to define scenario categories, as has been shown as a result of the CETRAN project~\cite{Gelder2020b}.
However, to the knowledge of the authors, no formal mechanism has yet been presented, on how a traceable and complete catalog can be derived.

\cite{WEB23} define the term \textit{scenario concept}, which defines scenario classes and puts them in relation with each other.
Furthermore, scenario concepts define a certain scope or an underlying assumption for the derivation of scenario concepts.
The benefit of scenario concepts is that a scenario catalog can be generated, which shows a clear line of reasoning for the derivation of the scenario classes.
\cite{Weber2019} provides a scenario concept that defines crash-relevant scenarios for the highway domain.
It derives scenarios around interaction with an object that challenges the ego-vehicle to avoid a collision.
Other traffic participants can take the role of contributing factors, by e.g. constraining the options of the ego vehicle for avoidance maneuvers.
\cite{WEB23} provide a holistic scenario concept for the definition of base-scenario, which covers bilateral interactions with other traffic participants.
To derive the base scenarios, reusable concepts are defined, which are used to distinguish scenario classes in different contexts, e.g. the concept \textit{following} can be used to define the scenarios \textit{following a leading vehicle} while in longitudinal traffic or following a vehicle while performing a left turn at an intersection.

\subsection{Completeness approaches}
\label{sec:completeness_approaches}

The completeness of a concept for scenario-based testing is a key issue to argue whether a function is safe. Within SOTIF therefore four quadrants are described distinguishing the Operational Design Domain (ODD) into known and unknown scenarios \cite{SOTIF21448}.
\cite{HAR72} thereby observes saturation effects in detected driving maneuvers.
\cite{GLA23b} approximate the degree of a certain set of scenarios by approximating the coverage of real-world parameter space coverage using a data-based bootstrapping approach.
Whereas coverage is normally assessed based on data, ontologies are used to aim for holistic descriptions of the environment. \cite{NEU21} created an ontology to explore potential influence factors and structure this along the 6 layer model \cite{Scholtes2021} to aim for a sufficient ODD description.
\cite{WEB23} uses an ontology to derive base scenarios from abstract traffic concepts to make sure to include relevant aspects.
Although there are ways investigated in literature to approach completeness, to the best of the authors' knowledge, nobody provides a sufficient argumentation for completeness.

\subsection{Argumentation techniques}
\label{sec:argumentation_techniques}

Arguing the safety of automated driving or sub-processes is crucial to ensure the safety of those systems. Therefore, literature towards the argumentation of complete systems and subcomponents exists for automated driving.
\cite{ISO15026} propose a goal structured notation (GSN) to systematically argue a safety case. A top claim is thereby argued using a strategy supported by subgoals, assumptions, and evidence.
This structure is e.g. used by \cite{WAR19} and \cite{RUD18}. \cite{WAR19} utilize a goal structured notation to argument pattern for multi-concern assurance of connected automated driving systems. \cite{RUD18} use a similar strategy for arguing AI safety in the automotive context. Thereby, large claims are structured and divided into subgoals in both papers.
\cite{FAV23} also argues the safety case and structures diverse claims along a claim-argument-evidence structure. Thereby, the evidences are graded along the confidence in the argument.
Although argumentation is performed on the system level, such argumentation is not yet performed for scenario-based testing to the best of the authors' knowledge.

\section{Methodology}
\label{sec:methodology}

To argue whether a scenario concept is complete or sufficiently complete, an argumentation structure is proposed to reduce the complexity of this problem. 
Therefore, relevant definitions are elaborated first.
Afterward, those are utilized along with known argumentation patterns to tackle the problem of completeness.
Lastly, strategies to gain evidence are elaborated to support the argumentation structure.

\subsection{Terminology}
\label{sec:definitions}

Terms such as completeness are defined in different ways in different domains and are even understood differently within certain contexts.
For a streamlined and unambiguous argumentation for scenario concepts, terms have to be clearly defined and have to be set into the context of scenario-based testing.
Therefore, definitions are proposed for the most important terms used in the paper:

\begin{itemize}
    \item \textbf{Scenario concept}: A scenario concept structures traffic into a set of predefined scenario categories within a scope and with a purpose. The concept contains scenario categories, their definitions, and relations between scenario categories \cite{WEB23}.
    \item \textbf{Completeness}: A scenario concept is sufficiently complete for a use case if all relevant driving situations are adequately captured. Thereby, the state of completeness is binary.
    \item \textbf{Coverage}: Coverage is the quantifiable extent to which a set of scenarios or parameters represent a defined ODD or predefined set of scenarios. 
\end{itemize} 

\subsection{Completeness and Coverage}
\label{sec:completeness_coverage}

\begin{figure}[tb]
\vspace{0.06in}
\centering
\includegraphics[width=\linewidth]{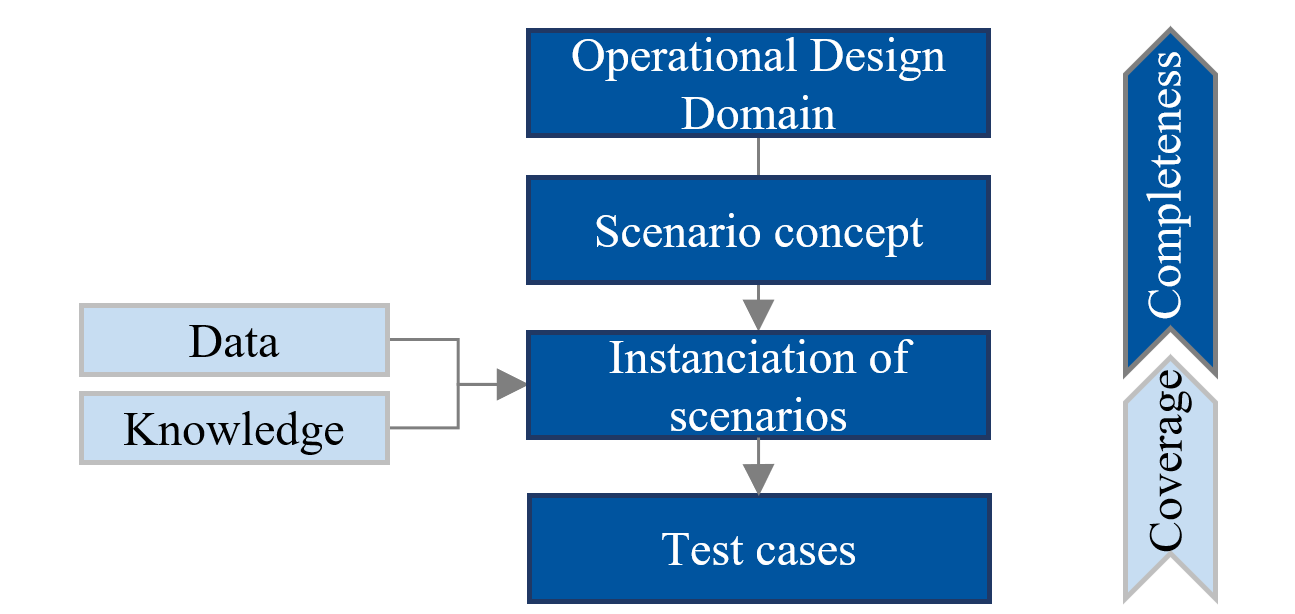}
\caption{Application of completeness and coverage}
\label{fig:coverage_completeness}
\end{figure}

The split between completeness and coverage leads to different fields of applicability for both terms in scenario-based testing (see Fig.~\ref{fig:coverage_completeness}).
The term completeness directly refers to a use case including a dedicated ODD and system under test or at least a rough ODD and a generic system under test depending on the specifications within the given use case. The systematic description of that ODD can therefore be sufficiently complete to evaluate a system under test. The term completeness is therefore practically not applicable to concrete instances within continuous spaces since it is impossible to describe all occurring concrete scenarios. Anyhow, it is possible to evaluate the concept of whether all important aspects are covered in the predefined logical scenario classes.

For the assignment of values to logical scenario classes, the term coverage is used. With a given scenario concept, values can be assigned from real-world data or knowledge to fill logical scenario classes with distributions. Those included data can cover a given space (either real-world or entities of a database) to a certain extent. So, claims as "x percent of scenarios are covered" are possible. This representation can close the gap between the concept and the real-world.

Based on the elaborated definitions two questions have to be raised for scenario-based testing:
\begin{itemize}
    \item Is a chosen scenario concept sufficiently complete for a use case and an ODD?
    \item Do the values ranges and distributions cover the ODD sufficiently?
\end{itemize}

\subsection{Argumentation Structure}
\label{sec:argumentation_structure}

\begin{figure}[tb]
\vspace{0.06in}
\centering
\includegraphics[width=\linewidth]{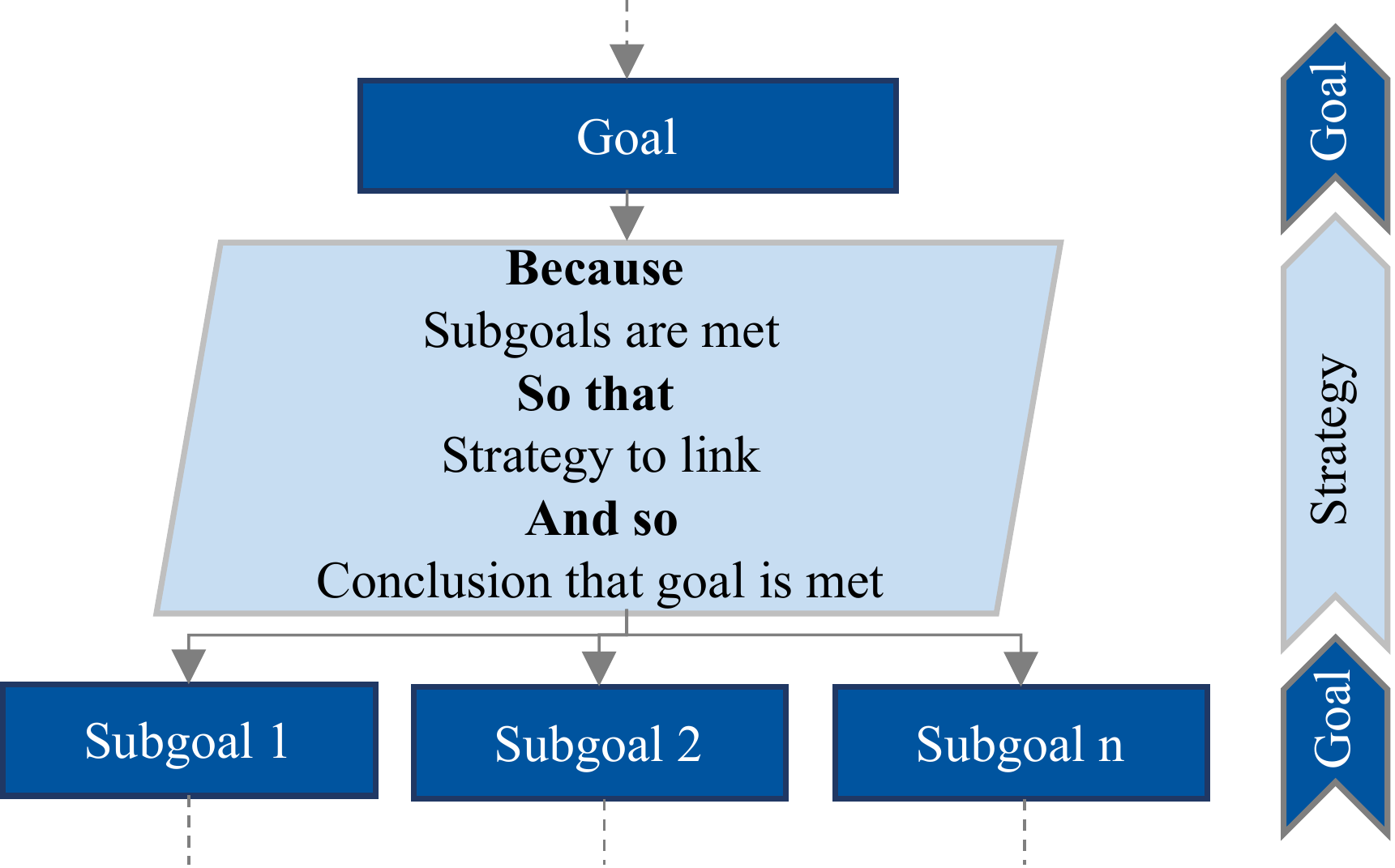}
\caption{Argumentation strategy \cite{TinoBrade23}}
\label{fig:gsn_strategy}
\end{figure}

In order to be able to answer both questions from the previous section, an argumentation structure is needed that can reduce the complexity of the questions. For this purpose, a GSN structure is used according to \cite{ISO15026}. 
In this structure, the high-level goal "scenario concept is complete" is deconstructed into individual parts.
To argue this goal, the concept must be defined completely and unambiguously and the ODD must be covered.
For the argumentation, the claims are divided into subclaims using strategies (see Fig.~\ref{fig:gsn_strategy}).
These are highly formalized to keep the argumentation process as error-free as possible and to provide a uniform structure for stakeholders.
This structure consists of an argument regarding the subgoals under the assumption that they are correct, a strategy to connect subgoals with the goals, and a conclusion.
The strategy is therefore only valid for the goal if recursively argued subgoals are also correctly argued.

The proof is essential for testing the strategy under the assumption that the subgoals are correct. 
However, this is not done utilizing direct proof, but by testing the counter hypothesis.
If this is confirmed, the strategy is inadequate, or the concept is not sufficiently complete.
If it is refuted by evidence, the strategy is confirmed.
In this way, all individual elements can be broken down and systematically scrutinized. This also makes it easier to identify shortcomings.

In addition to the argumentation of individual elements, the second component of the argumentation structure consists of high-level concerns that can be raised by different stakeholder groups. Relevant stakeholder groups are, for example, regulatory entities, simulation engineers, customers, or society.
Concerns derived from these are also decomposed after collection and checked using the initial argumentation structure.

\subsection{Strategies for Evidences}
\label{sec:evidences}

To prove individual counter hypotheses from Sec.~\ref{sec:argumentation_structure} right or wrong, evidence has to be given to support the argumentation structure and scrutinize the strategies. 
Evidence can come from different sources depending on the hypothesis to prove and the structure of the given scenario concept.
Therefore, strategies can be proven both indirectly or directly.
Whereas the actual evidence has to fit the strategy and has to be adjusted to it, a few common strategies are listed in Tab.~\ref{tab:evidences}: 
\begin{table}[]
\vspace{0.06in}
\caption{Evidence strategies to prove or refute arguments}
\begin{tabular}{|l|l|}
\hline
\textbf{Type}  & \textbf{Evidence strategy}   \\ \hline \hline
\begin{tabular}[c]{@{}l@{}}knowledge-\\ based\end{tabular} & \begin{tabular}[c]{@{}l@{}}- Plausibilization\\ - Expert surveys\\ - Atomicity principle\\ - Constraints of the system under test\\ - ...\end{tabular}  \\ \hline
data-driven  & \begin{tabular}[c]{@{}l@{}}- Detection/ analysis of real-world situations\\ - Observation of saturation effects\\ - Parameter sensitivity analysis in simulation results\\ - ...\end{tabular} \\ \hline
\end{tabular}
\label{tab:evidences}
\end{table}

Both, data-driven and knowledge-based evidence can be used to scrutinize different aspects of the concept.
On the one hand, constraints of the system under test may be well known e.g. the inability of a radar to detect colors. So, the absence of colors of entities would not lead to an incomplete concept for the given function.
On the other hand, it may be harder to identify all dynamic constellations and distributions between objects a priori. Therefore, real-world data should be analyzed.
So, the type of evidence needed highly depends on the strategy and needed confidence for a stakeholder group.

\section{Application}
\label{sec:application}

To prove the applicability of the proposed methodology, an exemplary completeness argumentation is shown. Therefore, an existing scenario concept is taken to fill the goal structured notation. Furthermore, specific methods for counterarguments are applied utilizing existing data.

\subsection{Scenario concept}
\label{sec:app_scenario_concept}

\begin{figure}[tb]
\vspace{0.06in}
\centering
\includegraphics[width=\linewidth]{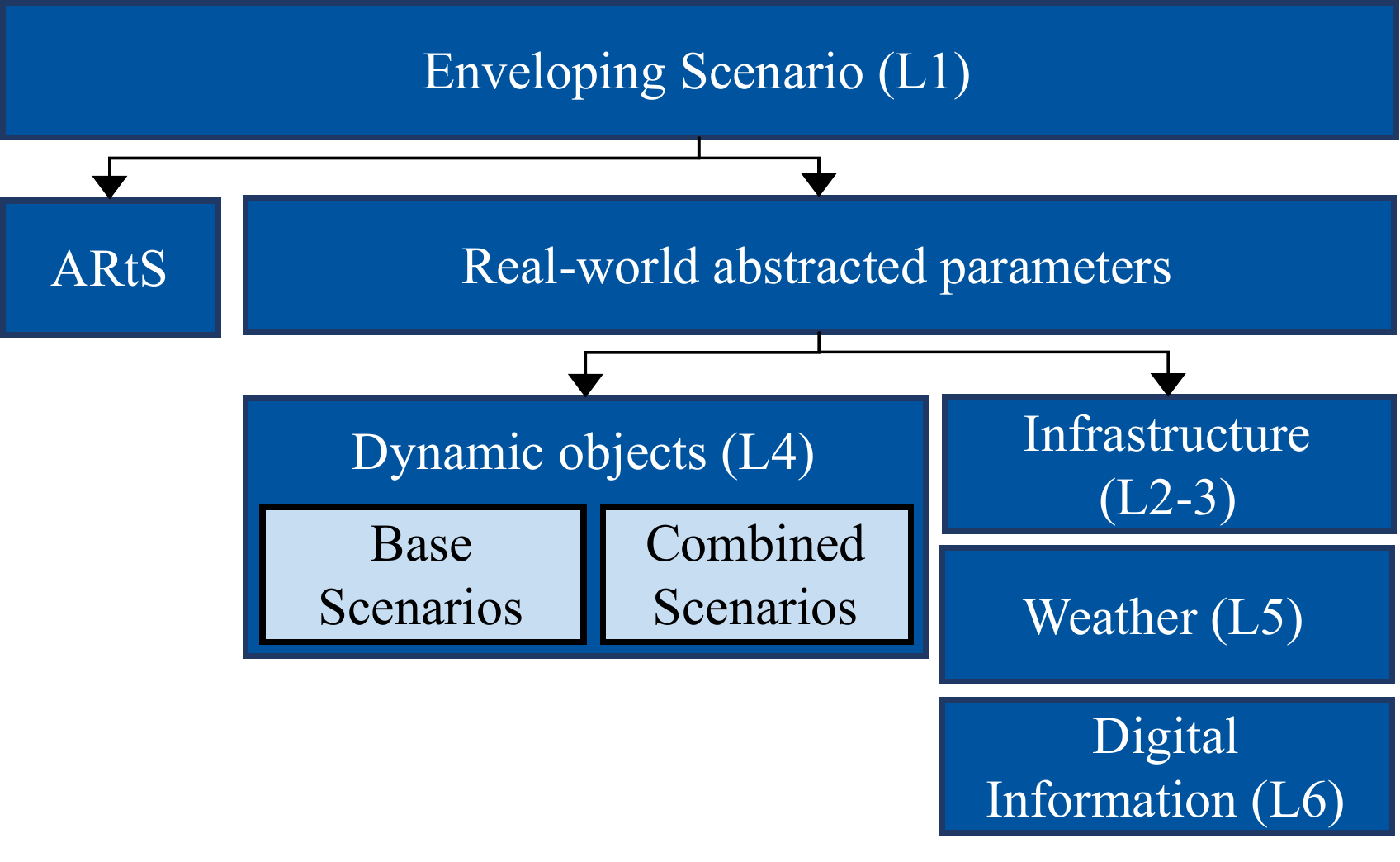}
\caption{Scenario concept structure}
\label{fig:scenario_concept}
\end{figure}

The completeness argumentation is applied to an exemplary scenario concept. Although multiple different concepts exist (see Sec.~\ref{sec:scenario_concepts}), the concept is set up in a hierarchical structure to allow a straightforward argumentation and to improve the explainability of the concept (see Fig.~\ref{fig:scenario_concept}).
To cut longer traffic sequences into small scenarios, \textit{enveloping scenarios} are defined first.
An enveloping scenario is thereby defined as a spatial and temporal limited canvas in which the scenario occurs. 
Whereby the spatial region is set by meaningful connected infrastructure elements, the temporal cut is defined by the enter and exit of the ego vehicle within this defined region. 
Using that concept, a longer ride of an ego road user can be cut into different intersections, roundabouts, and longitudinal streets.
Based on those enveloping scenarios, elements of other layers can be described in two ways: using specific parameters or an advanced replay similar to \cite{Weber2021a} recorded from real traffic. Whereby aspects of high interest for the use case should be modeled using specific parameters for its description, more aspects may belong to the scenario, but should not be modeled in detail to prevent a scenario space explosion. 
Therefore, a replay recorded from real traffic is used, but the surrounding traffic is adapted according to potential deviations from the observed behavior of the system under test in the recorded file.
For explicit scenario parametrization, parameters are structured according to the 6 layer model \cite{Scholtes2021} since they can be defined mostly independently.
Special attention is paid to layer 4 (dynamic objects) due to its complexity. Those dynamic objects and their relations are described using base scenarios \cite{WEB23} and focus scenarios. Whereas base scenarios are elementary descriptions between the ego and another object, focus scenarios can be adaptions or combinations of base scenarios to investigate special aspects of a scenario deeply or to combine road users to build more complex scenarios.

\subsection{Application to argumentation}
\label{sec:app_argumentation}

\begin{figure*}[tb]
\vspace{0.06in}
\centering
\includegraphics[width=\linewidth]{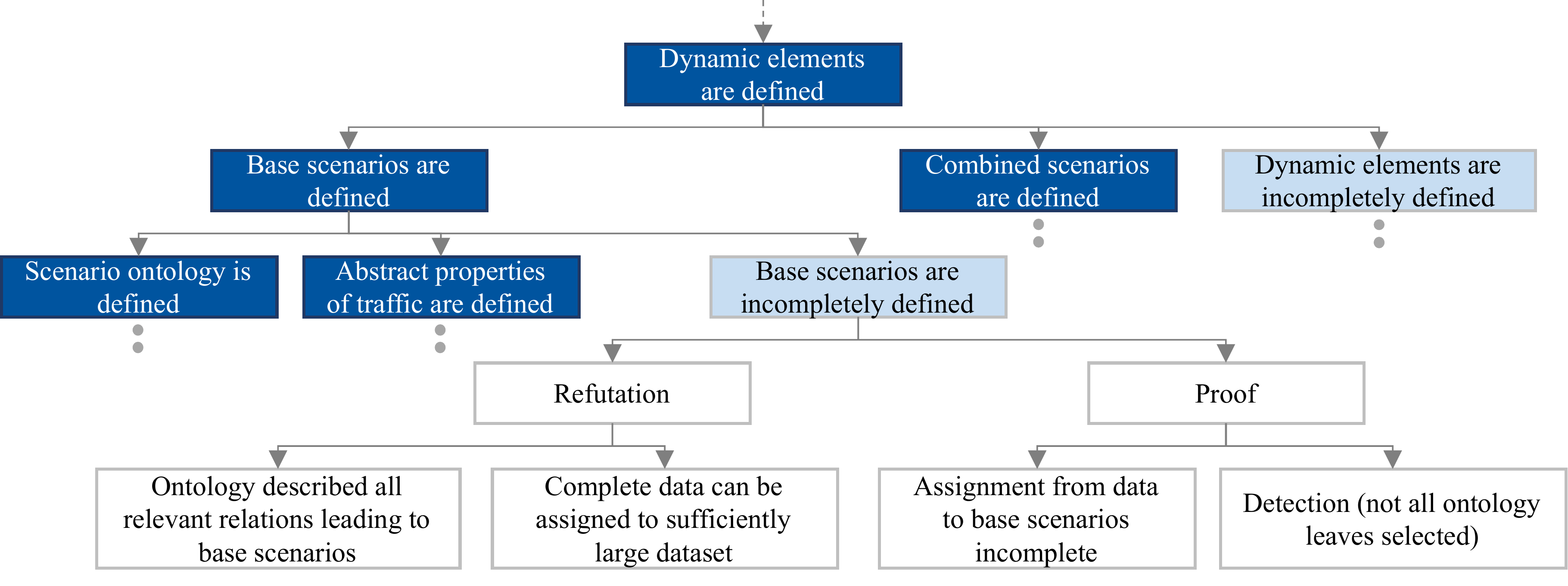}
\caption{Part of argumentation structure with goals (dark blue), counter hypothesis (light blue) and evidence strategies (white)}
\label{fig:example_argumentation_structure}
\end{figure*}

The GSN structure presented (see Sec.~\ref{sec:argumentation_structure}) is applied in full to the scenario concept presented. 
To limit the completeness argumentation process to a dedicated function within a use case, for this paper, a generic system under test is defined. 
It is assumed that a system under test only considers trajectories movements of dynamic objects as relevant for its behavior. Perception phenomena are also not considered in the following. 
Starting from the top goal that the scenario concept is complete, subgoals are decomposed and the structure of the concept is utilized. This is based on the levels of the 6 layer model (see Sec.~\ref{sec:app_scenario_concept}) as it is the state of the art that all elements of a scenario can be described in those layers \cite{Scholtes2021}.
Furthermore, the spatial and temporal split along enveloping scenarios is a simplification of the real-world. However, it is claimed that this split is sufficient since all spatial regions are covered by individual enveloping scenarios and only the time the ego vehicle moves in it is relevant. The claim that an enveloping scenario is sufficiently complete in itself has to be argued separately.
Therefore, the complete graph can be found on github\footnote[1]{https://github.com/ika-rwth-aachen/scenario-completeness-gsn}. 
An excerpt from layer 4 is presented below (see Fig.~\ref{fig:example_argumentation_structure}). 
According to the concept, the constellations are described by base scenarios and their combinations (combined scenarios).
Since the strategy says that elementary constellations should be captured by base scenarios this has to be proven or the counter hypothesis has to be refutated.
Knowledge-based and data-driven evidence are investigated for this.

Under the condition that abstract properties of the scenario concept describe scenarios sufficiently complete, an incompleteness of base scenarios can be refuted by the fact that all meaningful possible combinations of those abstract properties are derived using an ontology that is set adequately according to another subgoal.
The properties are thereby structured mutually exclusive so that properties describing similar aspects are grouped by a common concept which is used for the derivation of base scenarios.
The completeness of such group properties as well as further proof for the completeness of base scenarios is then given utilizing real-world data.

\subsection{Data-driven Evidences}
\label{sec:app_real_world_data}

The utilization of real-world data to prove whether a strategy holds true is a main pillar of the completeness argumentation structure. Whereas pure argumentation can help to find inconsistencies and scenarios that have been specified and are plausible but have not been observed in data yet, unforeseen scenarios cannot be identified with argumentation, but with external input.
Thereby, data cannot only be used to prove certain completeness claims but also to assess the coverage.

To apply evidence for the completeness of base scenarios and abstract properties as described in Sec.~\ref{sec:app_argumentation}, the inD dataset \cite{inD} is used. It consists of 13.499 trajectories from four urban intersections. Thereby it is far away from completeness for a reasonable ODD, but strategies can be shown exemplarily.

To prove data-driven that base scenarios are complete, real-world data is observed and base scenarios are assigned to all situations.
Therefore, rules are set to detect each base scenario type and its timespan according to \cite{VVM_D13}. 
Rules have to be formulated positively to avoid categorizing elements in base scenarios which are unknown. E.g. a categorization in \textit{vehicles} and \textit{non-vehicles} would not serve this condition unless it can be proven that this categorization would be sufficient for a system.
Applying the rules on the dataset, 59,253 base scenarios are found without a second in the recordings where no base scenario is assigned. 
So, the base scenarios are complete on the given abstraction level with regard to the inD dataset. 

Going into the subgoal that abstract properties have to be complete for the completeness of base scenarios, this completeness has to be proven as well.
Therefore, concrete parameters are assigned to those properties and it is checked whether the real-world detected base scenarios can be reconstructed with the defined parameters. Thereby, small deviations between real-world and reconstructed trajectories are seen in the evaluation of base scenarios in \cite{Weber2023}. 
Since only a generic system under test is used for this evaluation, whether this deviation is acceptable cannot be argued. Therefore, simulations would be needed.

\begin{figure}[tb]
	\centering
	\subfloat[Base scenario types]{
		\includegraphics[width=0.45\linewidth]{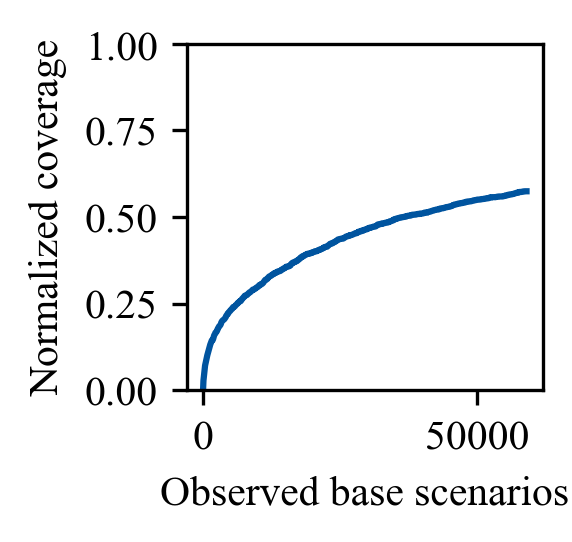}
		\label{fig:base_scenario_coverage}
	}
	\hspace{0.005\linewidth}
	\subfloat[Object start velocity in scenarios]{
		\includegraphics[width=0.45\linewidth]{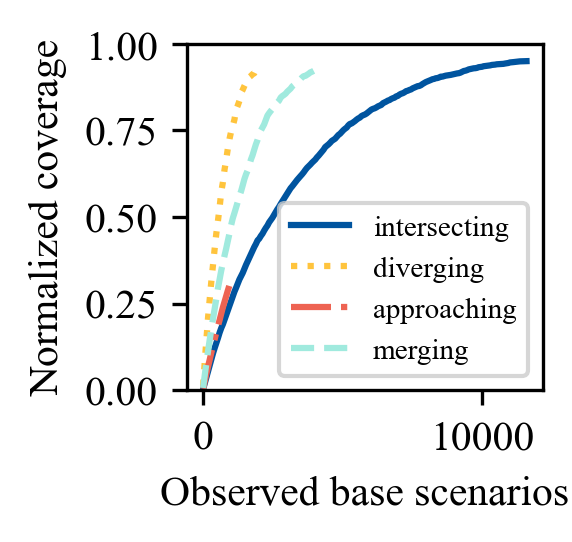}
		\label{fig:velocity_coverage}
	}
	\caption{Saturation of parameters in inD data \cite{inD}}
	\label{fig:combination_of_base_scenarios}
\end{figure}

Besides the completeness of the concept, data can also be utilized to assess the coverage of a set of scenarios as elaborated in \cite{GLA23b}. The coverage of data can thereby be assessed by approximating saturation functions on different abstraction levels.
For the given scenario concept, thereby the coverage of abstract scenarios and the coverage of parameter spaces in real-world data can be distinguished to complete the picture and argue that logical scenario instances are complete.
Extracting parameters from real-world data, a clear saturation behavior can be observed for both, the detection of different base scenarios (see Fig.~\ref{fig:base_scenario_coverage}) and individual parameters within these (see Fig.~\ref{fig:velocity_coverage}).
Thereby, it can be seen that different abstraction layers and parameters need different amounts of data depending on the underlying distributions.
To argue whether the saturation is sufficiently complete, occurrence probabilities of scenarios and parameters as well as confidence requirements are needed specific to the overall safety argumentation. 

\section{Discussion}
\label{sec:discussion}

Within section \ref{sec:application} the proposed methodology is applied to the dynamic definition of a scenario concept exemplarily.
In addition to the use of qualitative evidence to argue the model structure, quantitative evidence in particular plays an essential role. This is particularly important when the link to a larger argumentation structure of the overall safety case must be included, such as the link between coverage probabilities of the concept and a risk balance of the system under test.
The use of coverage and completeness considerations depends on the chosen level of abstraction of the concept.

Another separate point is the applicability of a scenario concept for simulations or other use cases. Completeness is therefore not a sufficient criterion for use, as it must e.g. also be possible to transfer it to suitable test environments and explain it to relevant stakeholders.

Even if the reasoning gives the impression that a concept is sufficiently complete at the time of application, this does not necessarily apply to later points in time. The reasoning is necessarily based on a state of knowledge of data or knowledge of traffic. However, this can change, e.g. due to new findings, new road user types or topologies, so that a regular comparison is necessary and, if necessary, updates must be made within the concept within a dev-ops process.

\section{Conclusion}
\label{sec:conclusion}
This paper presents a strategy for arguing the completeness of scenario concepts.
A distinction is made between completeness and coverage and a goal structured notation structure and evidence strategies are proposed for comprehensive argumentation. The argumentation is systematically scrutinized on an individual level and with top-level concerns. Furthermore, it is applied to an existing scenario concept and a publicly accessible real data set with a focus on dynamic objects and constellations.
A more detailed application on other layers and evidence for those remain future work.